\documentclass[preprint,showpacs,preprintnumbers,amsmath,amssymb,nofootinbib]{revtex4}

\usepackage{graphicx}
\usepackage{dcolumn}
\usepackage{bm}

\begin{document}

\preprint{SPECTRUM.TEX v13}

\title{%
   Comparison of the Ultra-High Energy Cosmic Ray Flux\\
   Observed by AGASA, HiRes and Auger
}%

\author{B.M.~Connolly}\email{connolly@nevis.columbia.edu}
\author{S.Y.~BenZvi}
\author{C.B.~Finley}
\author{A.C.~O'Neill}
\author{S.~Westerhoff}
\affiliation{%
   Department of Physics\\
   Columbia University and Nevis Laboratories\\
   New York, NY 10027
}%

\date{\today}

\begin{abstract}
The current measurements of the cosmic ray energy spectrum at ultra-high energies ($\text{E}>10^{19}$\,
eV) are characterized by large systematic errors and poor statistics.  In addition, the
experimental results of the two experiments with the largest published data sets, AGASA and
HiRes, appear to be inconsistent with each other, with AGASA seeing an unabated continuation
of the energy spectrum even at energies beyond the GZK cutoff energy at $10^{19.6}$\,eV.  Given
the importance of the related astrophysical questions regarding the unknown origin of these
highly energetic particles, it is crucial that the extent to which these measurements
disagree be well understood.  Here we evaluate the consistency of the two measurements for
the first time with a model-independent method that accounts for the large statistical and
systematic errors of current measurements.  We further compare the AGASA and HiRes
spectra with the recently presented Auger spectrum.  The method directly compares two
measurements, bypassing the introduction of theoretical models for the shape of the energy
spectrum.  The inconsistency between the observations is expressed in terms of a Bayes 
Factor, a standard statistic defined as the ratio of a separate parent 
source hypothesis to a single 
parent source
hypothesis.
Application to the data shows that the two-parent
hypothesis is disfavored.  We expand the method to allow comparisons between an experimental flux and that
predicted by any model.
\end{abstract}

\maketitle

\section{Introduction}
By measuring the energy spectrum of cosmic rays above $10^{19}$\,eV we 
hope to shed some light on the yet unknown sites and acceleration 
mechanisms that can produce a particle of such energies.  These measurements 
are intrinsically difficult, as the cosmic ray flux at ultra-high energies 
is low and we rely on earthbound detectors which cannot directly detect 
the primary particle.  Rather, its properties are reconstructed by 
measuring the secondary particles of the extensive air shower produced 
when the primary enters Earth's atmosphere. Low statistics and large 
systematic errors in the energy determination are typical for these 
measurements, which should consequently be treated with care.

The existing measurements of the energy spectrum above $10^{19}$\,eV have 
received much attention.  A longstanding hypothesis predicts that the flux 
of the highest energy cosmic rays should be suppressed above 
$10^{19.6}$\,eV as cosmic rays from distant sources will interact with the 
cosmic microwave background via photopion production until their energy 
drops below this threshold energy \cite{greisen,zatsepin}.  This so-called GZK suppression 
is in itself not a controversial prediction, but it has not been 
detected unambiguously, and several cosmic rays with substantially higher 
energies have been observed with various detectors over the years.

The most recent results disfavoring the GZK hypothesis come from the Akeno 
Giant Air Shower Array (AGASA) collaboration, whose published energy 
spectrum shows no indication of a high-energy suppression \cite{sasaki,agasa}.  In 
contrast, the monocular-mode energy spectra measured by the High 
Resolution Fly's Eye detectors (HiRes 1 and HiRes 2) support the existence 
of a GZK feature \cite{hiresI,hiresII}.  The current world data set is still 
small.  The AGASA claim that the GZK supression is not observed is based
on only 11 events above $10^{20}$\,eV.  In the near future,
the Pierre Auger Observatory, currently under construction in
Malargue, Argentina, will dramatically increase the world data set.  
The Auger collaboration has published a first energy spectrum based on 
1.5 years of data taken during construction \cite{auger1}.

A second problem is that although 
typically not shown in plots of the energy spectrum, the errors 
on the energy determination are also large: 30\,\% at $3\times 
10^{19}$\,eV and 25\,\% above $10^{20}$\,eV in AGASA \cite{agasa}, 
30\,\% in HiRes \cite{hiresI,hiresII} and 50\,\% in Auger \cite{auger1}. 
Furthermore, a correct evaluation of the systematic 
errors is difficult.  
It should for example be noted that the systematic uncertainty for the AGASA
energy scale includes a 10\,\% systematic uncertainty due
to the hadronic model.  This uncertainty is calculated 
by comparing different hadronic event generators and 
defining the systematic uncertainty by their difference.
This may not be a reliable estimate of this uncertainty, 
which could potentially be much larger and is, in any case,
unknown at this point.
Since the AGASA and HiRes experiments use radically different techniques to determine the 
primary energy -- AGASA is a ground array and HiRes an air fluorescence 
detector -- the errors have fundamentally different sources.  
\footnote{It should be noted that the published Auger spectrum has an energy scale
that is calibrated with its fluorescence detector.  If simulations are used
to determine the shower energies from surface detector data alone, the
energies are systematically higher by at least 25%
\cite{auger1}.}  

The large statistical and systematic errors quoted by the experiments 
raise the question how much significance should be attached to the 
discrepancy between the HiRes and AGASA spectra.  Several authors have addressed this question.  
In \cite{demarco1,demarco2}, the 
experimental results are compared individually to model predictions of the 
shape of the energy spectrum at GZK energies.  Such an analysis 
requires assumptions on the nature of the cosmic ray sources, as the shape of 
the GZK suppression will of course depend on the source distribution.

Before one asks if any of the current measurements can be used as 
evidence for or against a specific model for the origin of cosmic rays, 
one needs to answer the question of whether or not the two experiments actually 
disagree at all, given their uncertainties.  This requires a method to 
compare the two spectra that is both model-independent and considers the 
uncertainties quoted by the experiments by accounting for the probability 
of any systematic shift in energy scale applied to the data.

This paper presents a general spectrum comparison technique that 
for the first time addresses these points. The technique naturally incorporates the Poisson 
errors in the observed fluxes, and can probabilistically treat systematic 
errors in the absolute energy scale.  The technique is developed in 
Section 2 and applied to the energy spectra of AGASA \cite{agasa},
HiRes 1 and HiRes 2 in monocular mode \cite{hiresI,hiresII} and Auger 
\cite{auger2} in Section 3.  Section 4 
summarizes the results and describes how the method can be expanded for 
testing all experimental measurements against a theoretical prediction.

\begin{figure}[t]
\includegraphics[width=0.7\textwidth]{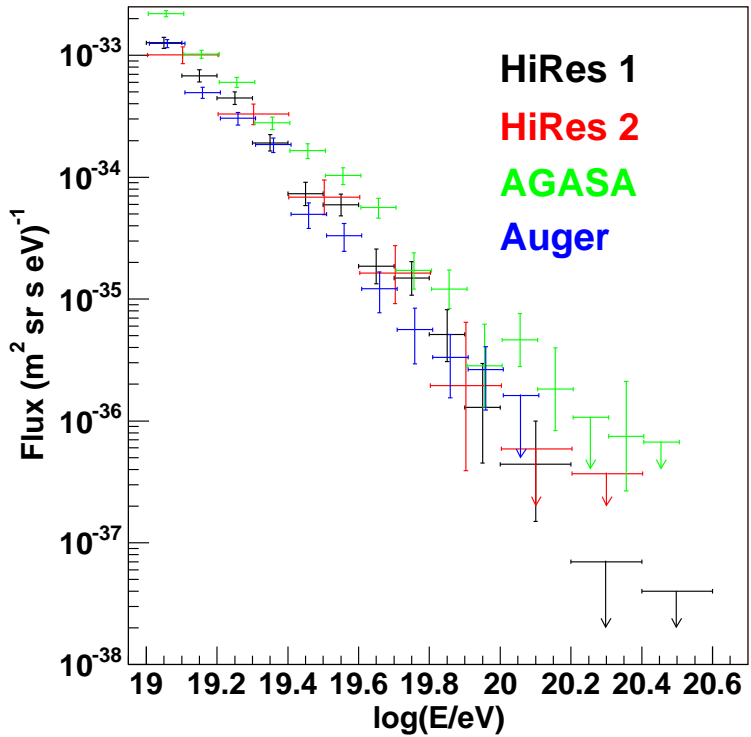}
 \caption{\label{fig:plot_0}  The AGASA \cite{agasa}, Auger \cite{auger2} and HiRes monocular spectra \cite{hiresI,hiresII}.  }
\end{figure}

\section{Proposed Comparison Between AGASA, HiRes and Auger}
The comparison between the AGASA and HiRes spectra can be quantified as the
relative probability that the various fluxes came from a signal parent
distribution versus two separate distributions.  This probability ratio is
quantified as a Bayes Factor ($BF$).  Two parent fluxes can arise for any
number of reasons, for example systematic mismeasurement by either
experiment, or -- however unlikely -- a different primary flux at AGASA or
HiRes.  

The calculation of the $BF$ is a variation of the method given in \cite{jeffreys}
comparing the consistency of two Poisson parameters.  The statistic 
is the ratio of the probability that the spectra arise
from separate parent distributions over the probability of
the alternative hypothesis that they are derived from a single parent
distribution.  The $BF$ then indicates to what degree the separate parent
hypothesis is favored.  A widely accepted interpretation  of the resulting
value for the $BF$ is shown in Table\,\ref{table:BF}.

To perform the calculation, we divide each energy distribution into $N$ bins.  For the
$i^{th}$ energy bin, the relative number of events is parametrized by the
variable $f_i$.  If the single-parent hypothesis is true, the relative
exposures of the two experiments should determine the number of events
expected from that single parent distribution.  In the case of the
two-parent hypothesis, a similar parameter, $f^\prime _i$, is invoked that can 
take any value for any bin and is not
constrained by the relative exposures of the experiments.  
The probabilities
of both hypotheses are calculated assuming Poisson uncertainties for the
fluxes in the $i^{th}$ bin, while marginalizing the energy scale
uncertainties.

\begin{table}[t]
\begin{small}
\caption{\label{table:BF}  The interpretation of the separate parent flux
hypothesis (SPFH) given the Bayes Factor ($BF$) \cite{jeffreys}.  \newline}
\begin{center}
\begin{tabular}{|c|c|}
\hline
 Bayes Factor ($BF$)    & Interpretation \\
\hline
  $BF>1 $  	                  & SPFH supported \\
\hline
  $10^{-\frac{1}{2}}<BF<1 $  	  & Minimal evidence against SPFH    \\
\hline
  $10^{-1}<BF<10^{-\frac{1}{2}} $ & Substantial evidence against SPFH \\
\hline 
  $10^{-2}<BF<10^{-1}$ 	          & Strong evidence against SPFH \\
\hline
  $BF<10^{-2}$ 	          & Decisive evidence against SPFH \\
\hline
  \end{tabular}  
\end{center}
\label{table:parameters}
\end{small}
\end{table}

We will derive the $BF$ using the comparison 
of the AGASA and HiRes spectra as an example;
however, in the end, we will calculate a separate $BF$ for every combination
of AGASA, HiRes 1, HiRes 2 and Auger spectra. 
The spectra for AGASA and HiRes at a given energy scale
are defined as ${A_i}$ and ${H_i}$, where $i \in [1,N]$.    
We define the percent shift in the AGASA and HiRes energy scales to
be $s_A$ and $s_H$, with the percent energy scale uncertainty
to be $\sigma_A$ and $\sigma_H$, respectively.  
Following the derivation in \cite{jeffreys}, 
the true number of events measured by AGASA and HiRes in
the $i^{th}$ energy bin is parameterized by
\begin{equation}
a_i = f_i \eta_i
\end{equation}
and 
\begin{equation}
h_i = (1- f_i) \eta_i,
\end{equation}
respectively, where
$\eta_i$ is the 
total number of hypothesized events expected for both the AGASA and HiRes
experiments.  For the hypothesis
where the two spectra come from a single parent distribution, 
\begin{equation}
\label{equation:f_i}
f_i= \frac{(AGASA~Exposure)_i}{(AGASA~Exposure)_i+(HiRes~Exposure)_i}.
\end{equation}

In the case where the AGASA and HiRes experiments are measuring separate 
parent distributions, we parametrize the hypothesized events in a similar
fashion.  For the separate parent hypothesis,
\begin{equation}
a^\prime _i = f^\prime _i \eta_i
\end{equation}
and 
\begin{equation}
h^\prime _i = (1- f^\prime _i) \eta_i,
\end{equation}
where $f^\prime _i$ is allowed to take any value and 
hence will be marginalized.

Making use of Bayes' Theorem, 
the Bayes Factor is 
\begin{eqnarray}
\label{equation:BF_pre}
BF &=& \frac{P(separate~parent~hypothesis|A,H)}{P(single~parent~hypothesis|A,H)}  \notag \\ 
   &=& \frac{\int \int P(A,H|f^\prime,\eta)q(f^\prime,\eta)df ^\prime d\eta }{\int P(A,H|f,\eta)q(f,\eta) d\eta} 
\end{eqnarray}
where, e.g., $A\equiv\{A_i\}$.
If one had a model with which to constrain $A$ and $H$, 
it would be included in the prior, $q(.)$.  Any unknowns in the model could then be marginalized
with $\eta_i$ and $f ^\prime _i$ provided that the priors on $\eta_i$ and $f ^\prime _i$ do not cancel.  
The denominator of Eq.\,(\ref{equation:BF_pre}) assumes some values for $f_i$ 
calculated through Eq.\,(\ref{equation:f_i}).  The numerator marginalizes $f ^\prime _i$
assuming any and all sets of $f_i^\prime$.  By marginalizing 
over $f_i^\prime$, we are effectively asserting that the relative exposure quoted 
by the two experiments (defined by $f_i$) does not hold.  Therefore, some adjustment needs 
to be made to one or both of the exposures.  Since we do not have any idea what that adjustment
could be, we sum over every possible relative exposure (i.e. marginalize $f_i^\prime$).  Note: if the 
difference in the spectra are due to some physical process, this would manifest itself in a change in
the relative exposures.

To introduce the energy scale uncertainties, we modify
\begin{align}
P(A,H|f,\eta)q(f,\eta) 
\rightarrow & P(A,H|f,\eta,s_A,s_H)\notag\\
&\times q(f,\eta,s_A,s_H)
\end{align}
to account for the deviations of the true energy scale from the 
mean energy scale of AGASA and HiRes, respectively.  $A$ and $H$
are made dependent on the energy scale, such that the values of $A_i$ 
and $H_i$ will depend on $s_A$ and $s_H$, respectively.\footnote[1]{Some may interpret this as changing the
measurements, $A_i$ and $H_i$, based on the choice of energy scale.  
We can escape this interpretation of ``changing measurements''
if we imagine that, as the various terms for $s_A$ and $s_H$ are calculated, those events
that fall inside the ``energy window'' between $10^{19.6}$ and $10^{21}$\,eV
are calculated according to Eq.\,(\ref{equation:BF}).  
The probability distribution for those events whose true energies
fall outside this window is flat ($\equiv 1$).  That is, 
we impose the arbitrary condition that if an event's true energy 
is $<10^{19.6}$\,eV or $>10^{21}$\,eV, then nothing can 
be said about the predicted number of events in that particular bin.  
In this way, we can calculate the $BF$ for a finite energy range,
avoid the concept of a changing measurement based on a changing hypothesis
and still shift the spectra relative to one another in a somewhat intuitive
manner.}  
If the numerator and denominator are properly evaluated in this form, 
they necessarily account for the probability of some observed fluxes, $A$ and $H$, given 
the energy scale.  As the deviations of the true energies 
from the measured energies are unknown, 
probability theory allows $s_A$ and $s_H$ to be marginalized 
such that Eq.\,(\ref{equation:BF_pre}) becomes
\begin{align}
BF =   \frac{\int \int \int \int P(A,H|f ^\prime,\eta,s_A,s_H)q(f ^\prime,\eta,s_A,s_H) df ^\prime d\eta ds_A ds_H}{\int \int \int P(A,H|f,\eta,s_A,s_H)q(f,\eta,s_A,s_H)d\eta ds_A ds_H}. \notag \\
\end{align}

The fluctuations in the fluxes are treated as Poisson, while the distributions
of $s_A$ and $s_H$ are parametrized by Gaussians.  Explicitly, the $BF$ is
\begin{align}
\label{equation:BF}
BF = & \frac{\int \int \left\{ \prod_{i=1}^{N} \left[ \int \int P_{f^\prime _i\eta_i}(A_i)  P_{(1-f^\prime _i)\eta_i}(H_i) df_i ^\prime d\eta_i \right] \right\} N(s_A;0,\sigma_A) N(s_H;0,\sigma_H) ds_A ds_H}
{\int \int \left\{ \prod_{i=1}^{N} \left[ \int P_{f_i\eta_i}(A_i)  P_{(1-f_i)\eta_i}(H_i) d\eta_i \right] \right\} N(s_A;0,\sigma_A) N(s_H;0,\sigma_H) ds_A ds_H } 
\end{align}  
where $P_{\mu}(M)$ denotes a Poisson distribution with mean $\mu$ and
$M$ measured events and $N(x;\mu,\sigma)$ denotes a Gaussian with 
a deviation, $x$, from the mean, $\mu$,
and error $\sigma$.
By integrating $\eta_i$ within the product, we are effectively summing
over every imaginable set of Poisson-distributed spectra.  
Each set of $A$ and $H$, where $A_i$ and $H_i$ are dependent on the
energy shift, is weighted by the probability of that 
shift.  
Since the size of the shift toward the true
energy scale is unknown, these shifts are marginalized by integrating  $s_A$ and $s_H$.
The numerator has an integration over $f^\prime _i$, effectively marginalizing over
the cases where $A$ and $H$ come from different parent distributions.

To get a sense of the behavior of the $BF$, we apply the method to a simple toy example.
We generate two distributions, $S$ and $B$,  each with 20 bins;
distribution $S$ has 200 events, and distribution $B$ has $\beta\times 200$ events
such that $f_i\equiv\frac{\beta}{1+\beta}$ where $f_i$ is identical for every bin.
In effect, $\beta$ defines the relative exposures of the two ``experiments'' $S$ and $B$.
The number of events in each bin of $S$ and $B$ fluctuates according to Poisson statistics.

\begin{figure}[t]
\includegraphics[width=0.7\textwidth]{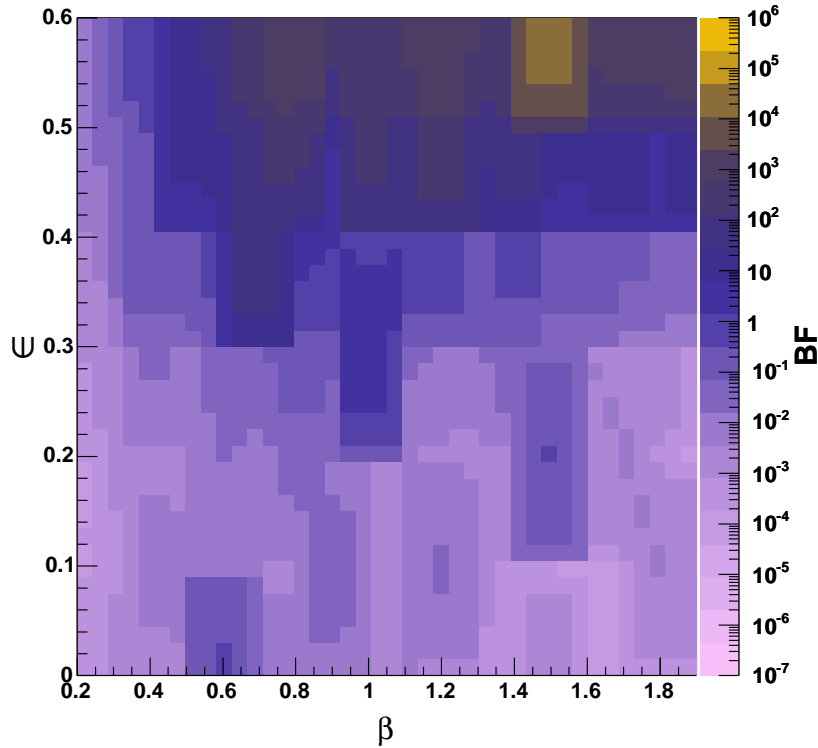}
 \caption{\label{fig:epsilon_f} $BF$ as a function of $\epsilon$ (the difference between the
 two parent distributions) and $\beta$ where the $BF$ is evaluated over
 two spectra with 200 and $200\times\beta$ events.}
\end{figure}

On top of these ``statistical'' fluctuations, we introduce a second fluctuation of 
distribution $B$, while the mean of the $S$ distribution is kept flat.    
Each bin of the distribution $B$ is evenly fluctuated within 
$\pm\epsilon$, constraining the total events in 
$B$ to $200\times\beta$, thereby making the 
distributions more different as $\epsilon$ increases. 
Therefore, $\epsilon$ quantifies the disagreement between $S$ and $B$.
Note that in contrast to the statistical fluctuations, these $\epsilon$
fluctuations are not included in the $BF$ and 
therefore deviations between $S$ and $B$ due to such fluctuations
are  `perceived' by
the $BF$ as differences in the spectra.
We then plot the $BF$ as a function of $\beta$ and $\epsilon$ in Fig.\,\ref{fig:epsilon_f}. 

The $BF$ distribution, given $\epsilon$ and
$f$, has substantial tails that
cause large fluctuations in Fig.\,\ref{fig:epsilon_f} 
even when averaged over hundreds 
of thousands of trials.
Despite these fluctutations, we are still able to pick out 
trends in the $BF$ as a function of $\beta$ and $\epsilon$.  
For instance, the $BF$ systematically increases with $\epsilon$ (i.e. as the difference
between the distributions increases).  
As $\beta$ increases, the statistics for $B$ increase.  
Therefore, it is more likely that a given difference between two distributions
is due to a real difference in the parents rather than low statistics. 

\section{Implementation}
Due to practical limitations for the calculation of
Eq.\,(\ref{equation:BF}), some choices must be made in the binning and
shifting of the data. The calculation occurs over 14 bins of size
$\log{E/\mathrm{eV}}=0.1$ from $10^{19.6}$\,eV to $10^{21.0}$\,eV.  The
energy bins are determined from existing data.  The $10^{19.6}$\,eV cut-off
is motivated by a previous comparison in \cite{demarco1}.  The AGASA
experiment has 866 events above $10^{19}$\,eV and 72 above $10^{19.6}$\,eV.
HiRes 1 and HiRes 2 have 403 and 95 events above $10^{19}$\,eV and 35 and 6
events above $10^{19.6}$\,eV, respectively.  Auger has 444 events
above $10^{19}$\,eV and 17 above $10^{19.6}$\,eV.  

After integrating over $f ^\prime _i$ and $\eta_i$, Eq.\,(\ref{equation:BF}) reduces to 
\begin{align} BF = \frac
{\int \int \left[ \prod_{i=1}^{N} \frac{1}{A_i+H_i+1} \right] N(s_A;0,\sigma_A)N(s_H;0,\sigma_H) ds_A ds_H }
{\int \int \left[ \prod_{i=1}^{N} f_i^{A_i}(1-f_i)^{H_i}
\frac{(A_i+H_i)!}{A_i!H_i!}  \right] N(s_A;0,\sigma_A)N(s_H;0,\sigma_H) ds_A ds_H}
\end{align}

Since one cannot split a spectrum into an infinite number of bins,
one must make finite shifts in the two energy spectra.  The chosen shifts
are made $\pm 5 \sigma$ from the mean energy, and are incremented in
$\log{E/\mathrm{eV}} = 0.1$ steps, weighted by the Gaussians integrated over
the energy bin.  As various 
energy scales $s_A$ and $s_H$, are evaluated, the fluxes in the 14 bins 
change.  The shifts $s_A$ and $s_H$ are evaluated down (up) to $\log{E/\mathrm{eV}}=-0.5$ ($\log{E/\mathrm{eV}}=0.5$). 
For instance, energy bins below $10^{19.6}$\,eV shift in and out of the 
set of 14 bins depending on the values of $s_A$ and $s_H$.  
In the case of the 
HiRes 2 spectrum where the data are quoted in increments of 
$\log{E/\mathrm{eV}} = 0.2$, the events are divided in proportion to the absolute size
of the $\log{E/\mathrm{eV}} = 0.1$ bin.  Meanwhile, the relative exposures of AGASA and HiRes ($f$) 
remain unchanged.

\begin{figure}[t]
\includegraphics[width=0.7\textwidth]{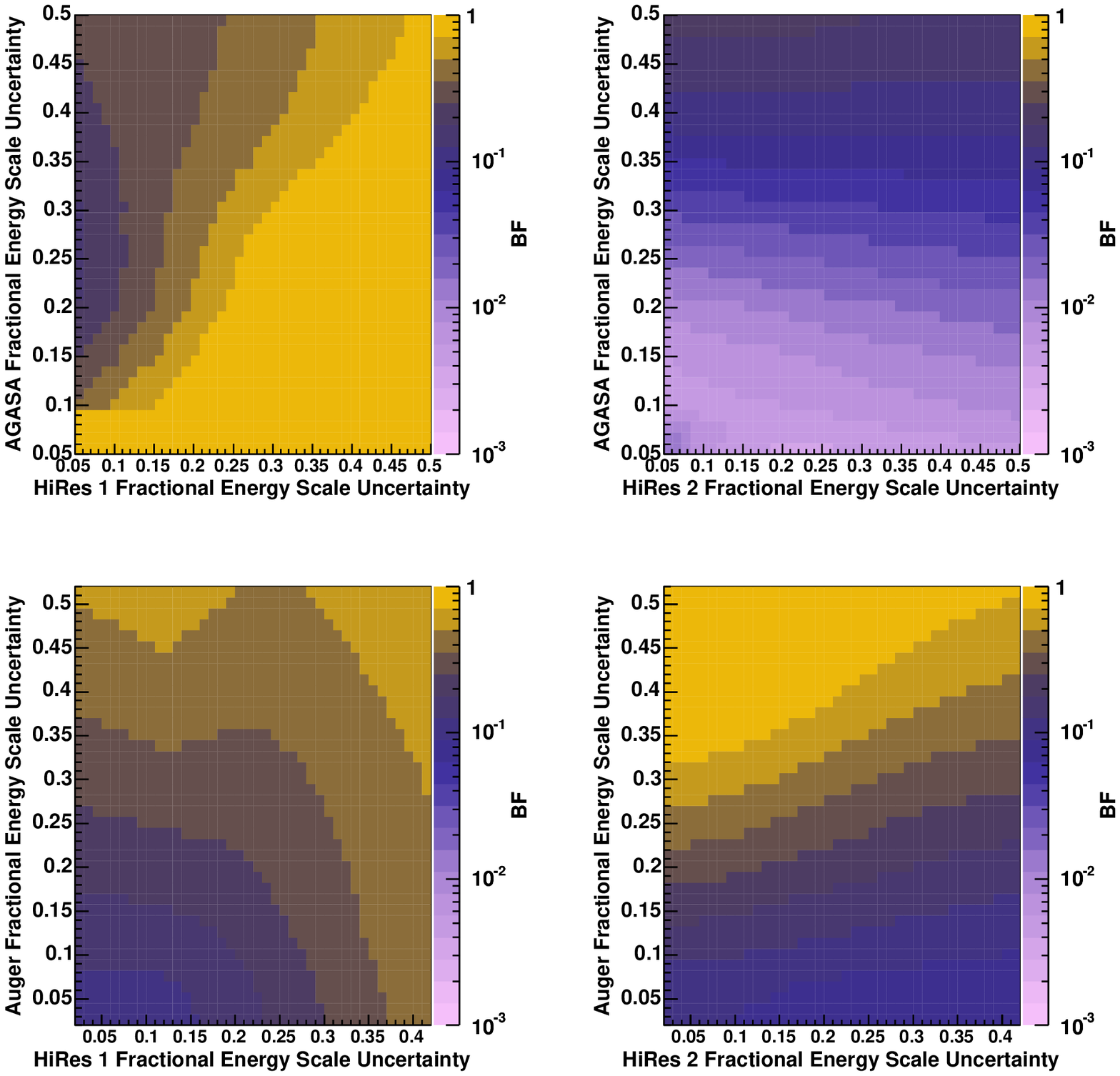}\\
\includegraphics[width=0.34\textwidth]{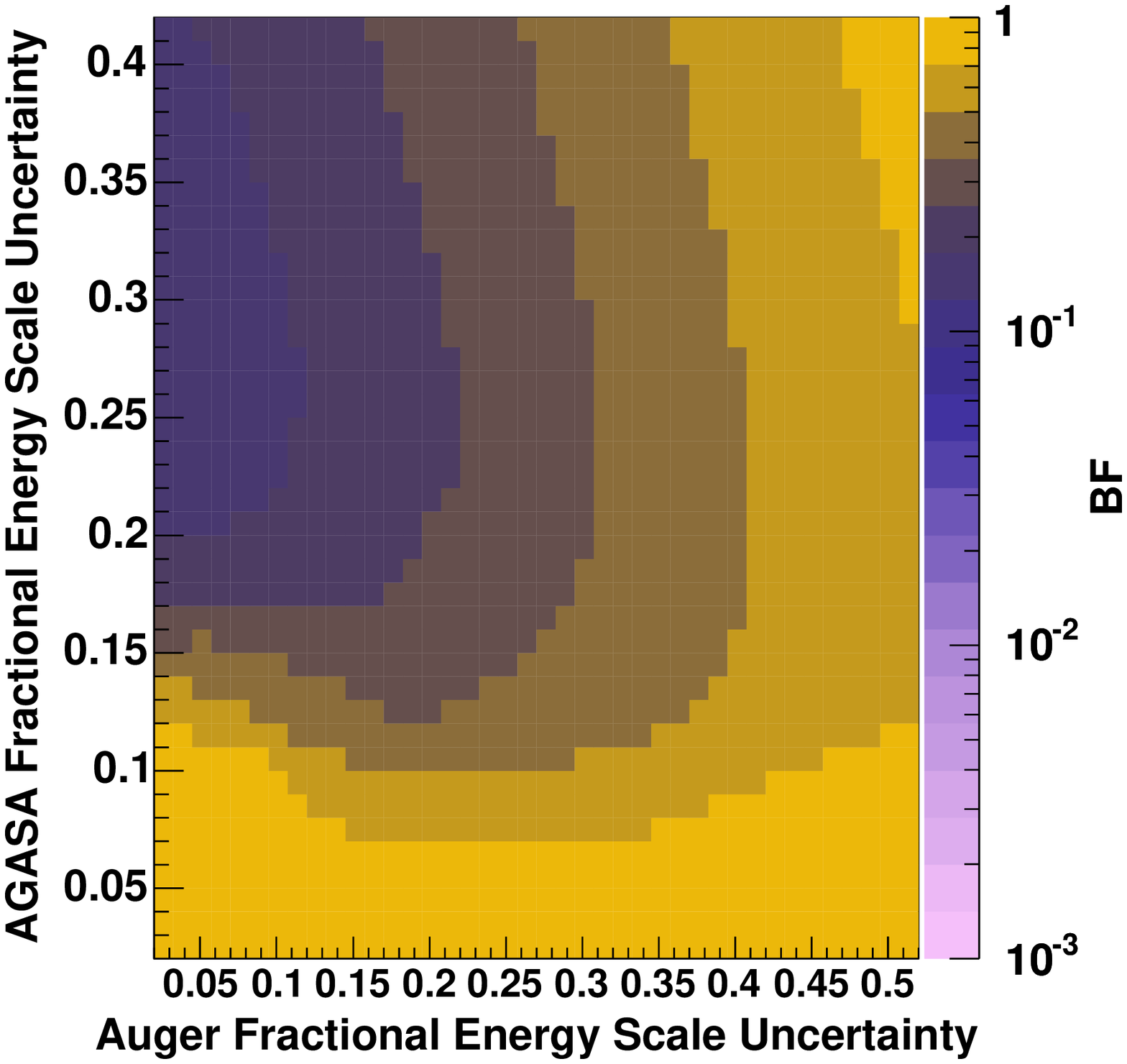}
 \caption{\label{fig:BF}  The Bayes Factor ($BF$) as a function 
 of the percent energy scale uncertainties for the compared spectra.
}
\end{figure}

Fig.\,\ref{fig:BF} shows 
the $BF$ as a function of the percent energy uncertainties for
all combinations of AGASA, HiRes 1, HiRes 2 and Auger spectra.
The distributions are non-trivial to understand as changing the 
energy scale does not amount to a simple shift in the number of 
events from one energy bin to another.  Rather, in HiRes, 
the shape of the spectra change as the energy scale changes because 
the exposure remains fixed as a function of energy.  
Further, the probability that the spectra agree for a given 
energy scale is convoluted with the probability that that energy scale 
is the true energy scale for the detector.  Meanwhile, 
the lower energy bins fall in and out of the calculation 
as the energy scales change in the course of the calculation.  

If we compare AGASA and HiRes with $30\%$ energy scale uncertainties
for both experiments, the
$BF=0.71$ and $BF=0.04$ for HiRes 1 and HiRes 2, respectively.
According to Table\,\ref{table:BF}, the $BF$ for the HiRes 1 comparison corresponds to
minimal evidence against the hypothesis that 
AGASA and HiRes are derived from separate distributions.  
The latter value for HiRes 2 corresponds to strong evidence against
the separate source hypothesis.  If we compare the Auger
spectrum with 50\,\% energy scale uncertainties
with the HiRes spectra with 30\,\% energy scale uncertainties, the 
$BF=0.54$ and $BF=0.85$ for HiRes 1 and HiRes 2, respectively,
corresponding to minimal evidence against the separate 
source hypotheses (i.e. no support for the two-source hypothesis).  
Finally, if we compare 
the Auger spectrum with 50\,\% energy scale uncertainties
with the AGASA spectrum with $30\%$ energy scale uncertainties
we obtain a $BF=0.74$ again corresponding to minimal evidence
against the separate source hypothesis.

In order to verify that the $BF$ we obtain from the experimental data is indeed consistent with
the single source hypothesis, we compare the actual $BF$ for the spectra from experiment $X$ and $Y$ to
simulations of the $BF$ generated under the assumption that they are generated from a single parent.  

\begin{figure}[t]
\includegraphics[width=1.1\textwidth]{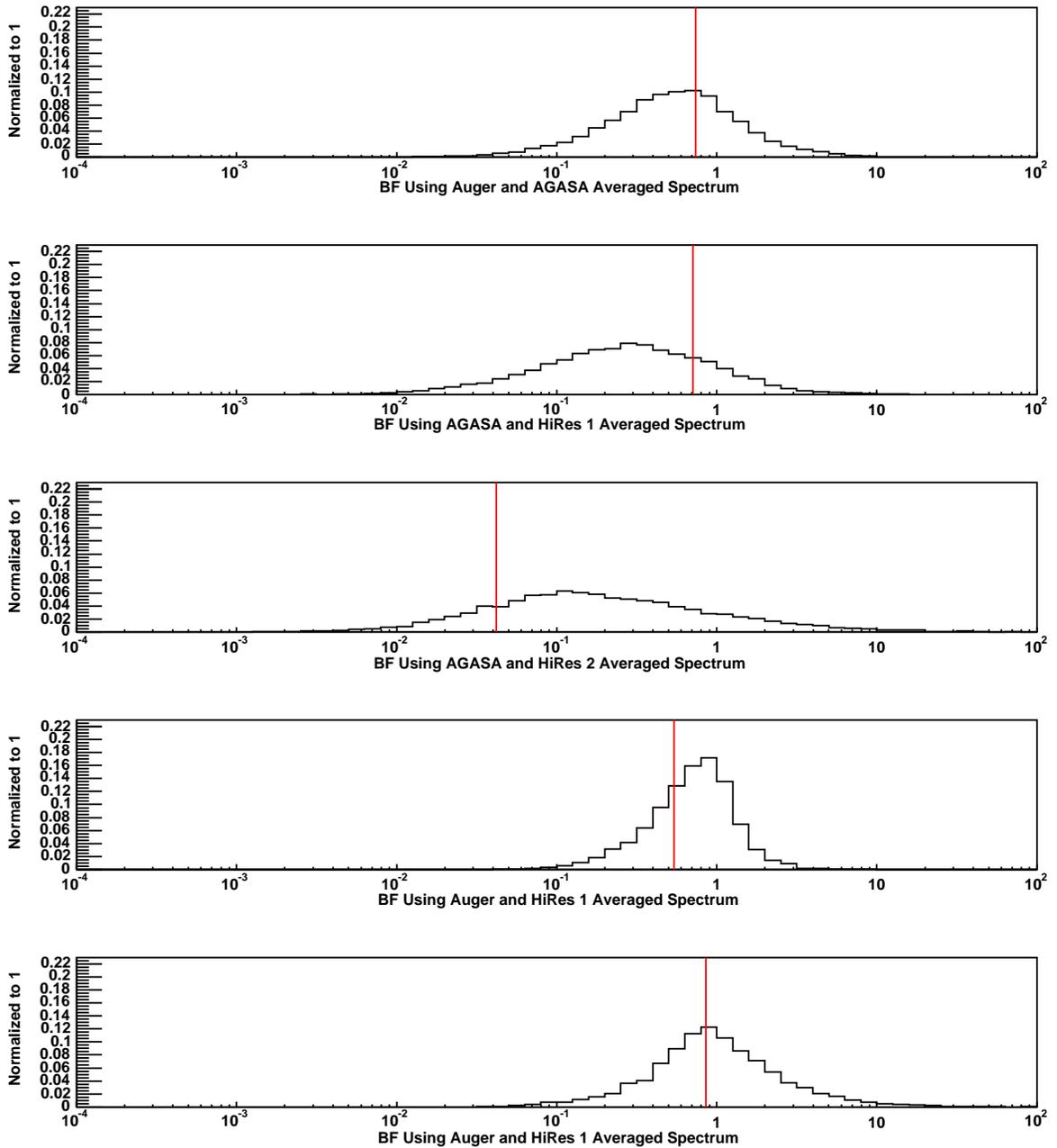}
 \caption{\label{fig:R_distributions_ave}  The distribution of the $BF$ calculated
 for an averaged spectrum from two experiments. 
 The red lines denote the $BF$'s measured for data. }
\end{figure}

To
calculate a simulated $BF$ for the spectra from experiments $X$ and $Y$, we assume that the true spectrum
is a spectrum that is an average, weighted by the exposure, of the spectra measured from
experiments $X$ and $Y$.  This average spectrum is then scaled to the 
corresponding exposures for
$X$ and $Y$.  We then randomly fluctuate both spectra according to
Poisson statistics and calculate the $BF$.  This procedure is repeated many times.
The resulting distribution of the $BF$ for two spectra generated assuming a single parent is then
compared to the experimental value of the $BF$ for the two measured spectra.  Fig.\,\ref{fig:R_distributions_ave} 
shows the results for a comparison of all three experiments with each other.

Comparing the $BF$'s measured for data (denoted by a red line) 
and those calculated in Fig.\,\ref{fig:R_distributions_ave}, we see that, in general, there is a 
substantial probability of obtaining a $BF$ that is larger than the 
$BF$'s calculated for the data.  That is, a substantial
number of samples that are derived from a single 
source look more like they were derived from two sources
than the real AGASA, HiRes and Auger comparisons.  
The value of the $BF$ tells us that the hypothesis that the AGASA and HiRes energy spectra 
are consistent with a single distribution is favored over the hypothesis that
they are derived from separate distributions.  

\section{Discussion}
\label{discussion}
 This result is similar to
those obtained for the AGASA and HiRes spectra in \cite{demarco1,demarco2}
by a rather different method.  Our method directly compares two
experimental results rather than comparing each to a theoretical prediction.  It also
treats systematic errors correctly by weighting possible systematic shifts in the energy
distribution by the probability of that shift. 

This result does not say anything about what the specific (in)consistencies would be of
either experiment given a theory.
However, our results indicate that there is room for a theory 
that agrees with each pair of spectra.
As the statistical power of the world data improves, the methods
discussed here will provide a convenient framework to compare experimental results with 
one another and with theoretical models.    

\begin{figure}[t]
\includegraphics[width=0.7\textwidth]{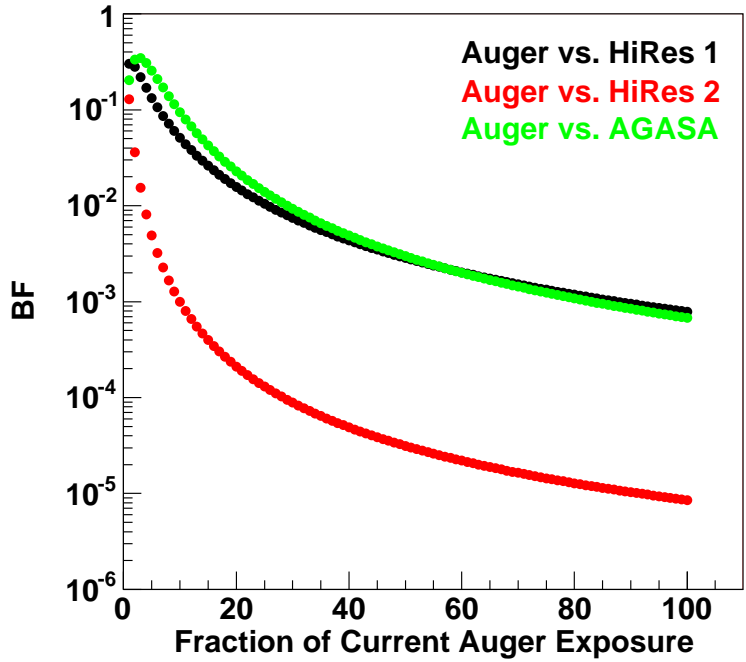}
 \caption{\label{fig:projected}  The projected $BF$ between
 the current Auger spectrum and the AGASA and HiRes spectra 
 as a function of the fraction of the current Auger exposure. 
 The calculations assume 
 30\,\% energy uncertainties in AGASA and HiRes and the projected
 15\,\% energy uncertainties in the Auger spectrum.
}
\end{figure}

The Auger experiment is currently the only operating ultrahigh energy cosmic ray
detector.  The Auger data set will increase dramatically over the next years.
Figure\,\ref{fig:projected} shows the projected value of the $BF$ between Auger
and the other experiments as a function of 
the fraction of the current Auger exposure.  
The $BF$ is calculated assuming the current 
spectra, the existing
30\,\% energy uncertainties in AGASA and HiRes and the projected
15\,\% energy uncertainties in the Auger spectrum.
In the limit where the exposure for Auger becomes large, 
the $BF$ decreases (increasing support for the one-source hypothesis). 
That is, the statistical and energy scale uncertainties in HiRes and AGASA are 
so large that, barring a significant change in the Auger spectrum, one will
never be able to tell whether or not the Auger and AGASA (HiRes) spectra are
derived from different parent distributions provided that we limit our analysis
to events above $10^{19.6}$\,eV. 
Note, however, that if one fits a model to one of the experimental 
spectra, our result does not imply consistency of the model with the
other experiment.  
\footnote{In order to compare a model with both
experiments, one must compare the model to both data sets simultaneously.
The above method can be modified to provide a statistic 
analogous to a $\chi ^2$ probability that could 
measure the consistency of theory given all experimental results.
In this case,
Bayes' theorem is used to obtain
\begin{equation}
P(\hat{t}|A,H)=\frac{P(A,H|\hat{t})q(\hat{t})}{\sum_{All~theories} P(A,H|t)q(t)}. 
\end{equation}
where, because of the definition of $f$ ($f^\prime$), each
$\hat{t}_i$ is the total number of events expected in the $i^{th}$ bin
for both experiments,
and, likewise, each $t_i$ is the number of events expected 
from any physically meaningful theory.  
Using the experimental Poisson and Gaussian errors,
\begin{align}
P(\hat{t}|A,H)= 
\frac{\int \int \left[ \prod_{i=1}^{N} \frac{f_i^{A_i}(1-f_i)^{H_i}\hat{t}_i^{A_i+H_i}e^{-\hat{t}_i}}{A_i!H_i!} \right] N(s_A;0,\sigma_A)N(s_H;0,\sigma_H) ds_A ds_H}
{\int \int \left[ \prod_{i=1}^{N} \frac{1}{A_i+H_i+1} \right] N(s_A;0,\sigma_A)N(s_H;0,\sigma_H) ds_A ds_H} \notag \\
\end{align}
where all theories are effectively marginalized by integrating $t_i$ within the product.
To obtain a $\chi ^2$-like probability,
one would calculate $P(\hat{t}|A,H)$, and then calculate many
$P(t^\prime|A,H)$'s where each $t^\prime_i$ is the Poisson fluctuated $\hat{t}_i$.
The fraction of events where $P(t^\prime_i|A,H)<P(\hat{t}_i|A,H)$ would
be the degree of confidence in the theory.  
}

This project is supported by the National Science Foundation under contract
NSF-PHY-0500492.

\end{document}